\documentclass[twocolumn,prl,preprintnumbers]{revtex4-1}
\usepackage{amsmath,amssymb}
\usepackage{verbatim,graphicx}

\def\tr{\textrm{tr}}

\begin{document}
\setlength{\unitlength}{1mm}

\preprint{MAD-TH-11-04}

\title{Evidence for fast thermalization in the plane-wave matrix model}
\author{Curtis T. Asplund$^a$, David Berenstein$^{a,b}$, Diego Trancanelli$^c$  \\
{\em $^a$ Department of Physics, University of California at Santa Barbara, CA 93106\\
$^b$ School of Natural Sciences,  Institute for Advanced Study, Princeton, NJ 08540\\
$^c$ Department of Physics, University of Wisconsin, Madison, WI 53706}}

\begin{abstract}
\noindent
We report on a numerical simulation of the classical evolution of the plane-wave matrix model with semiclassical initial conditions. Some of these initial conditions thermalize and are dual to a black hole forming from the collision of D-branes in the plane-wave geometry. In particular, we consider a large fuzzy sphere (a D2-brane) plus a single eigenvalue (a D0-particle) going exactly through the center of the fuzzy sphere and aimed to intersect it. 
Including quantum fluctuations of the off-diagonal modes in the initial conditions, with sufficient kinetic energy the configuration collapses to a small size.  We also find evidence for fast thermalization: rapidly decaying autocorrelation functions at late times with respect to the natural time scale of the system.
\end{abstract}
\maketitle

\paragraph*{-- Introduction and conclusion.}
From the beginning of the gauge/gravity correspondence \cite{M} it was understood that large black holes with anti de Sitter asymptotics should be related to thermal states in the dual field theory \cite{Witten1}. So, formation of a black hole from non-thermal initial conditions should be dual to thermalization of a specific initial state in the dual dynamics. This idea has led to the suggestion that the rapid thermalization observed in heavy ion collisions \cite{rhic} should be described by a dual black hole formation event \cite{Nastase}. 

Nascent black holes settle very quickly to the no-hair solutions, as shown by numerical simulations (see, {\it e.g.}, \cite{eg}). 
We expect that the dual theory will behave similarly. It has also been conjectured that black holes are ``fast scramblers'' \cite{SS}, \emph{i.e.},  they distribute information (or wash it away) faster than any other physical system, logarithmically in the number of degrees of freedom. This refers to how fast small quantum fluctuations away from equilibrium settle back to equilibrium in a black hole. 

The purpose of this paper is to explore such thermalization processes from the point of view of the field theory and to formulate a program where the fast scrambler conjecture can eventually be tested by numerical methods. We do this by focusing on a system with finitely many degrees of freedom where a collision of two gravitons or D-branes at high energy can in principle be studied: the plane-wave (or BMN) matrix model \cite{BMN}. This paper describes the first simulations of this system we have performed and the evidence we have acquired for fast thermalization. The simulations solve the classical equations of motion of the model. We show that the time averages of quantities in the system after thermalization match the Gibbs distribution for some degrees of freedom that appear quadratically in the Hamiltonian. The temperatures measured by various of these degrees of freedom are the same. We also show that various gauge invariant quantities have autocorrelation functions rapidly decaying in time with respect to the natural time scale in the system (defined by the data rather than machine time).

Ideally we would study this problem in the matrix model of \cite{BFSS}, which is dual to M-theory on flat space as the discrete light-cone formulation of the theory. In that system one has asymptotic states that can be scattered and could lead to a black hole formation event with an S-matrix interpretation. However, the initial conditions for that setup are not understood: the gravitons are bound states at threshold that necessitate solving the many body quantum dynamics in detail. Monte-Carlo simulations of the matrix dual black holes of this model have shown that the ensemble is unstable due to the flat directions of the matrix model potential \cite{CW}. The instability can be regulated with mass terms that are naturally present in the BMN model. Euclidean computations of the BMN model at equilibrium have been performed in \cite{CvA}. This paper deals with the time-dependent classical regime in the theory and the dynamics of thermalization.

The BMN matrix model, which represents the plane-wave geometry with maximal supersymmetry, solves the problem of describing gravitons by having a different classical solution for each graviton. Each of these gravitons is represented by a fuzzy sphere. We lose the ability to perform collisions with asymptotic states that scatter from each other.  However, as shown in \cite{Berenstein:2010bi}, there are initial conditions where the fuzzy spheres are displaced with respect to each other and we can set up periodic brane collisions (crossings) instead. The period for these setups, $\tau$, is independent of the details of the graviton branes and their velocities. This lets us measure time with a clock that has a well-defined physical interpretation. These crossings have classical instabilities associated with them: some degrees of freedom are tachyonic during a brief time of the $\tau$-period and grow exponentially in the number of crossings until the system back-reacts. The growth of fluctuations has been understood with a linearized analysis \cite{Berenstein:2010bi} and in the present letter we extend that analysis to the rest of the thermalization process where the dynamics is very non-linear. Classically these fluctuations can be set to zero, but they will be present in the full dynamics because of quantum mechanics. 

We add noise to represent quantum fluctuations in the initial conditions. Our configurations seed the modes that would become tachyonic during some portion of the $\tau$-period \cite{Berenstein:2010bi}.
We initialize each such mode with a gaussian probability function with a width determined by $\hbar$ and the adiabatic frequency of the mode. Once these fluctuations grow sufficiently they take over and scramble the system.  The time scale for the initial exponential growth is logarithmic in the size of the initial fluctuations, which are proportional to $\sqrt{\hbar}$. We consider the system to thermalize quickly if, subsequent to this period, the decay of fluctuations back to equilibrium is fast. The analysis we do is valid strictly only when $\hbar$ is small. The solutions have large energy of order $N^2$, where $N$ is the size of the matrices. The energy does not scale with $\hbar$. If these systems thermalize, their temperature is large in quantum units. This is exactly the regime where classical physics is valid, for we only have finitely many degrees of freedom and all of them will have large quantum numbers and can be treated classically (this sets the limit of how big we can make $\hbar$ in practice). This means in particular that we can ignore the fermions, for they only affect the low temperature regime.

In the rest of this letter we discuss the numerical implementation of the BMN matrix model and the evidence for fast thermalization of the initial conditions we chose. 

\paragraph*{-- Numerical implementation.}
The bosonic degrees of freedom of the BMN matrix model are the hermitian matrices $X^{i=0,1,2}$ and $Y^{a=1,\ldots , 6}$ and their canonical conjugates $P_i$ and $Q_a$. The bosonic part of the Hamiltonian is
\begin{eqnarray}
H&=& \frac 12\, \tr  \Big(P_i^2+ Q_a^2 +(X^i+i \epsilon^{ijk} X^j X^k)^2\nonumber\\
&&\hskip 1cm +\frac 14 (Y^a)^2- [X^i,Y^a]^2-\frac 12[Y^a,Y^b]^2\Big). \label{H}
\end{eqnarray} 
We have rescaled the variables so that the classical equations of motion are independent of $\hbar$ and all the quantum mechanics is hidden in the initial conditions. We have also normalized the mass of $X$ to one, \emph{i.e.}, we measure time by the oscillation period of one of the $X$ modes. 

Because of the $\mathrm{U}(N)$ gauge symmetry we must enforce the Gauss' law constraint:
\begin{equation}
C=  [X^i, P_i]+[Y^a,Q_a] =0\,. \nonumber
\end{equation}
To solve the equations of motion we use a leapfrog algorithm and we record the absolute value of the constraint $\tr(C^2)$ as a check for the code. We find that the constraint is well satisfied for the runs we perform, so we do not need to implement constraint damping.

The main sources of difficulty are the initial conditions. For this paper, we have used the following initial classical configuration:
\begin{eqnarray}
X^0 &=& \begin{pmatrix}L^0_{n} &0\\
0&0 \end{pmatrix}, \,  \,
X^1=\begin{pmatrix} L^1_{n} & \delta x_1\\
\delta x_1^\dagger&0\end{pmatrix}, \,   \,
X^2 = \begin{pmatrix} L^2_{n} & \delta x_2\\
\delta x_2^\dagger&0\end{pmatrix}, \nonumber \\
P^0&=&\begin{pmatrix}0 &0\\
0&v \end{pmatrix}, \quad P^{1,2}=0= Q^{1, \dots ,6}, \quad
Y^a = \delta y^a. \nonumber
\end{eqnarray}
The dimension of the matrices above is set to $N=n+1$. The $L^i$ are $\mathrm{SU}(2)$ angular momentum matrices in the $n$-dimensional representation. This is a {\it fuzzy sphere} of size $n$.
The system has an additional eigenvalue that is initially at the origin with velocity $v$ in the positive $X^0$ direction. These are the initial conditions discussed in \cite{Berenstein:2010bi} with the addition of fluctuation seeds $\delta x, \delta y$.  The $\delta x, \delta y$ are generated randomly using a complex gaussian distribution with a width proportional to $\sqrt {\hbar/n}$. We interpret these as quantum fluctuations of the off-diagonal degrees of freedom.

Recall that the ground state of an oscillator with Hamiltonian $2 H= p^2 + \omega^2 x^2$ has a gaussian wave function with squared width $\langle x^2 \rangle = \hbar/2\omega$. In our case, because of the initial conditions, all of the off-diagonal modes between the lone eigenvalue and the fuzzy sphere have approximately the same frequency of oscillation, proportional to  $n$ \cite{VRetal}. 

The $\delta x^\dagger$ are determined by forcing the matrices to be hermitian. All the off-diagonal $\delta y$ components are generated by the same gaussian distribution, whereas the diagonal ones are set to zero since they are subleading in $N$. This is a very rough approximation for the $Y$ modes connecting the fuzzy sphere to itself, lumping them together as if they all had the same mass. We do not add fluctuations in the modes connecting the fuzzy sphere to itself in the $X$ variables as the unstable modes grow so quickly that such fluctuations are not required. If the system thermalizes the fine details of the initial conditions get washed out at later times, so we only need them to be qualitatively correct. Notice that our initial conditions are built to exactly satisfy $C=0$ while preserving the typical size of quantum fluctuations for the space variables. This is why we have no fluctuations of the $P$, $Q$ variables nor of the off-diagonal modes of $X^0$ connecting the lone eigenvalue and the fuzzy sphere. 

The discretized matrix equations of motion read
\begin{eqnarray}
X_{t+\delta t}= X_t + P_{t+\frac{\delta t}{2}} \delta t \,, \qquad
P_{t+\frac{\delta t}{2}} = P_{t-\frac{\delta t}{2}} -\frac{\partial V}{\partial X}\Big |_t \delta t \,,\nonumber
\end{eqnarray}
and similarly for the $Y$ modes. Here $V$ is the potential obtained from eq.~(\ref{H}). The parameter $\delta t$ and the total number of iterations of the leapfrog algorithm can be varied in the numerical code.  We record the matrix configurations every few steps.

Due to the accumulation of numerical rounding errors, every few steps we need to force the matrices to be hermitian. We do this right before recording configurations. We also vary the random seed to generate an ensemble with gaussian distributions and check that the results are robust against these variations. A more detailed description of the code and a more comprehensive presentation of the numerical results and their interpretation will be presented elsewhere (also see supplement). 


\paragraph*{-- Results.}
We now describe some useful ways to visualize the information contained in the $X$, $Y$ modes and their time derivatives. To describe the thermodynamics, we need to coarse grain the degrees of freedom, which must be gauge invariant combinations of $X$ and $Y$. The simplest such combinations are traces of matrix products. We can use these to compare different values of $N$ and to study the large $N$ thermodynamic limit. Note that the traces of the matrices $X^i$ and $Y^a$ are decoupled: the nonlinear parts of the equations of motion are commutators, so the traces of the matrices evolve independently from the rest of the system. The trace of $X^i$ oscillates with angular frequency $\omega=1$, while the trace of $Y^a$ oscillates with $\omega=1/2$. Because of our initial conditions, only the trace of the $X^0$ mode is excited and it serves as our clock. 

\begin{figure}[ht]
\includegraphics[width=3.0 in]{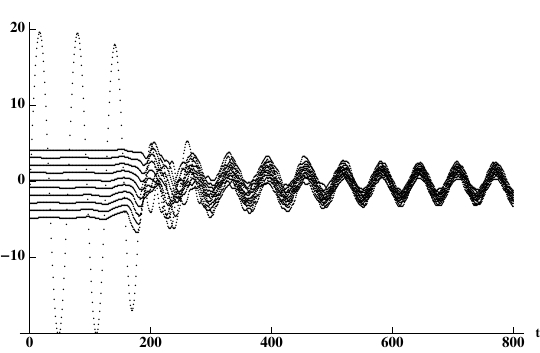}
\caption{Eigenvalue evolution for $X^0$. Here and in the following figures we have set $n=10$, $v=20$, and $\hbar=0.001$.  The time axis is in  discrete time units between recordings of configurations.}
\label{fig:eigenv}
\end{figure}

The other invariant way to work with matrices is in terms of their eigenvalues. These can tell us about the dual D-brane geometry. When the matrices are approximately commuting there is a clear geometric interpretation: the eigenvalues are positions of D-branes. When the matrices do not commute they still serve to roughly describe the distribution of D-branes inside the fuzzy object.  In Fig.~\ref{fig:eigenv} we plot the eigenvalues of $X^0(t)$.
Initially all of the motion is in the lone eigenvalue and the other eigenvalues are evenly spaced -- a property of the fuzzy sphere. As time goes by, the eigenvalues collapse to a much smaller vertical extent and oscillate collectively. The eigenvalues become very unevenly spaced and upon zooming in appear to repel each other, showing a typical behavior of random matrices. Qualitatively, they behave like a Dyson gas \cite{Dyson}, but a detailed comparison is beyond the scope of the present paper.

\begin{figure}[ht]
\includegraphics[width=3.0 in]{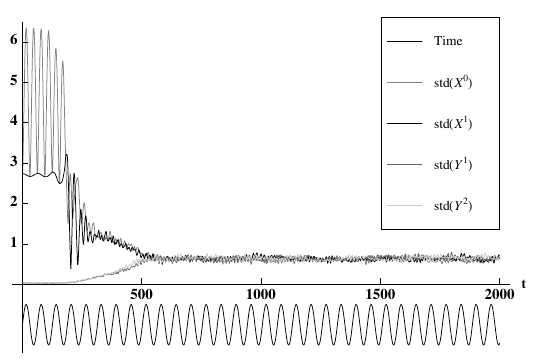}
\caption{Standard deviation of eigenvalues for the matrices $X^{0,1}$ and $Y^{1,2}$. We use the trace of $X^0$, rescaled,  to keep track of time (black curve at the bottom). Other values of $n, v,$ and $\hbar$ are qualitatively similar, see supplement.}
\label{fig:equil}
\end{figure}

It is also interesting to study the size of the system in different directions. We do this by evaluating the standard deviations of the eigenvalues of the matrices, see Fig.~\ref{fig:equil}. 
As shown in the figure, the fuzzy sphere collapses in size substantially. After the sphere has largely collapsed, the $Y$ modes grow from zero and converge to a value that is very close to the late-time value for the $X$ modes. Their growth is controlled by the random time variation of their effective mass after collapse. It is because of this that we needed to include fluctuations for most of the $Y$ modes. The subset of $Y$ modes connecting the fuzzy sphere and the eigenvalue do not grow enough in the initial phase and the time for fluctuations of those modes to converge is substantially longer. The size has only small fluctuations after convergence and the system seems to stabilize rapidly. The figure suggests that the object is becoming nearly spherical, a property shared by black holes without angular momentum. However, the corresponding dual black holes should have some deformation since they are not in asymptotically flat space.

\begin{figure}[ht]
\includegraphics[width=3.0 in]{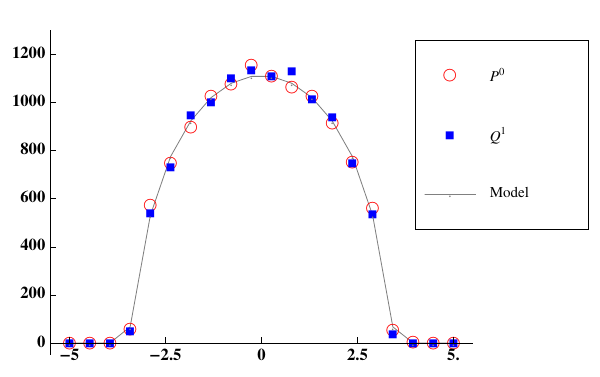}
\caption{The eigenvalues of $P^0$ and $Q^1$ are binned from $t=5000$ to $t=20000$ every ten steps, using the same time units of Fig.~\ref{fig:equil}. The semicircle distribution is integrated over the binning intervals and normalized to the total count of eigenvalues. The width is matched using the standard deviation of the eigenvalue distribution.}
\label{fig:semicircle}
\end{figure}

To test for thermalization, we compare time averaged distributions over successive configurations to those of the Gibbs ensemble for the classical system at some temperature $T$. Using the Gibbs measure $ dP\, dQ \exp(-H/T) $ we see that the momentum variables factorize into gaussian integrals. Thus the momenta are determined by the gaussian ensemble for hermitian matrices. It is well known that the distribution of eigenvalues (sufficiently coarse grained) should be a semicircle. We test this for the $P^0$ and $Q^1$ matrices starting way after the system looks thermalized ({\it e.g.}, after $t\sim 600$ in Fig.~\ref{fig:equil}). We wait until $t=5000$ to measure thermal properties just to make sure. This is shown in Fig.~\ref{fig:semicircle}.
The semicircle model matches the data well for both the $X$ and $Y$ momenta, which have the same distribution. This suggests that the system has thermalized, as the temperature measured from the $X$'s is the same as that measured from the $Y$'s.

\begin{figure}[ht]
\includegraphics[width=2.8in]{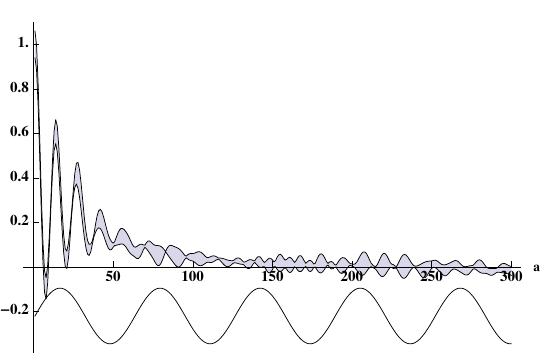} 
\includegraphics[width=2.8in]{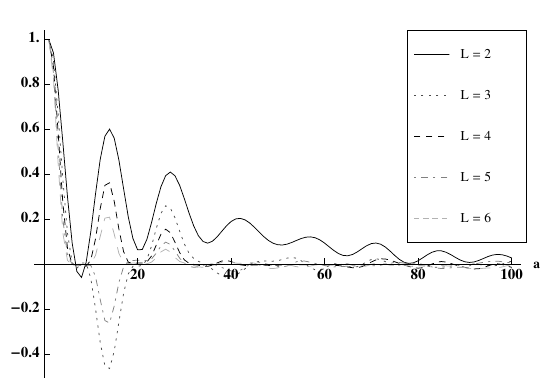}	
\caption{ Top: normalized autocorrelation of $\tr(X^1+iX^2)^2$. The width of the band indicates our statistical uncertainty from few similar sequences related to each other by rotations. We include our clock measure (black curve at the bottom). Bottom: we compare different $L$ modes for a shorter time period. The configurations are averaged in time starting at iteration 5000. 
Other values of $n, v,$ and $\hbar$ are qualitatively similar, see supplement.} 
\label{fig:autocorr1}
\end{figure}

Now that we have numerical evidence for thermalization we can study near-equilibrium configurations and fluctuation decay rates. This information can be obtained from the autocorrelation function  $\langle {\cal O}(t) {\cal O}^\dagger(t+a)\rangle$, where ${\cal O}(t)$ is some classical gauge invariant observable. 
This is an application of the fluctuation-dissipation theorem. This is averaged over $t$ well after thermalization. The simplest observables we can consider are of the form $\tr (X^1+iX^2)^L$. Similar traces are identified with graviton modes in ${\cal N}=4$ super Yang-Mills theory \cite{M}, where they are interpreted as having angular momentum $L$ along the dual $S^5$. Here, $L$ denotes angular momentum in the $\widehat{12}$ plane of the $X$ variables. Higher $L$ values correspond to higher spherical harmonics in the dual geometry. For $L=1$, as we have seen, the mode is decoupled, so the simplest non-trivial case will be for $L=2$, shown in Fig.~\ref{fig:autocorr1} (top). 
We can see that the autocorrelation for $L=2$ dies off quickly, with respect to the natural external clock. We also note that it oscillates in an interesting pattern, indicating that there are internal oscillation times of the variables associated with the thermalized system. Relative to these internal oscillations the autocorrelation function decays quickly (by half within 2 oscillations), so the associated vibration modes have a low quality factor. This indicates fast thermalization. In Fig.~\ref{fig:autocorr1} (bottom) we compare autocorrelations for higher $L$. We notice that the higher the $L$, the faster the autocorrelations decay. This is expected from black hole physics and the membrane paradigm of the horizon: when information approaches the membrane, it diffuses along the membrane until it becomes uniform. Diffusion happens first at short distance scales and then cascades to large scales. So this is evidence for an approximate notion of locality in the angular directions even in the thermal regime.
 
{\em -- Acknowledgements. } DB would like to thank F. Dyson, D. Kabat,  J. Maldacena, and H. Verlinde for discussions. 
Work supported by DE-FG02-91ER40618 (CA and DB) and DE-FG02-95ER40896 (DT). 

\newpage


\section{Supplement}

This is a supplement to ``Evidence for fast thermalization in the plane-wave matrix model", as appeared on {\it Physical Review Letters}. The supplement contains technical details on how  the data was extracted from the simulation, as well as how the data was reduced. We include additional plots on studies that show that the plots contained in the letter reflect the data broadly. We also include some additional discussion on the meaning of different thermalization times and what are the systematics that affects their numerical evaluation.


\subsection{Methodology }

\paragraph{-- Equations of motion and numerical parameter choices.}

Given the Hamiltonian
\begin{eqnarray}
H&=& \frac 12\, \tr  \Big(P_i^2+ Q_a^2 +(X^i+i \epsilon^{ijk} X^j X^k)^2
\cr&&\hskip 1cm 
+\frac 14 (Y^a)^2- [X^i,Y^a]^2-\frac 12[Y^a,Y^b]^2\Big)\,,\nonumber
\end{eqnarray} 
the equations of motion for the $X$ and $Y$ fields are different. They are given by
\begin{eqnarray}
&& \dot X^i = P_i\,,\cr
&& \dot Y^a = Q_a\,,\cr
&& \dot P_i = - X^i -3 i \sum_{jk}\epsilon^{ijk} [X^j,X^k]\cr && \hskip 1.4cm +\sum_k [[X^k,X^i],X^k] +\sum_a [[Y^a,X^i],Y^a] \,, \cr
&& \dot Q_a = -\frac 14 Y^a +\sum_i [[X^i,Y^a],X^i]+\sum_j [[Y^j,Y^a],Y^j]\,.\nonumber
\end{eqnarray}
Notice that for large  values of $X,Y$ and for $X,Y$ somewhat random the frequency of the modes is parametrized roughly by the absolute value of the eigenvalues of $X$ as a matrix.
Also notice that if we take traces we find that 
\begin{eqnarray}
&& \tr{\dot X^i} =\tr{P_i}\,,\cr
&& \tr{\dot Y^a }= \tr{Q_a}\,,\cr
&& \tr{\dot P_i} = -\tr{ X^i} \,, \cr
&& \tr{ \dot Q_a} = -\frac 14\tr{ Y^a }\,.\nonumber
\end{eqnarray}
These are closed sets of linear equations and they are harmonic oscillators of frequencies $1$ and $\frac12$ respectively. These give us our clock. The amplitude of $\tr{X}$ has a periodic motion and this period is independent of the details of the initial condition. Checking that this mode has this periodic motion serves as a test of the code.

From this information the time steps between iterations on our code should be controlled by the maximum size of $X$. For our simulations we have that $X$ is of order 10 or 20, so the time steps $\delta t$ should be such that $\delta t\ll 2\pi /20\simeq 0.3$ so that a single oscillation of the fastest mode is covered by many points.
For the particular simulation that is quoted in the letter we choose $\delta t = 0.004$ and we save the configurations every $25$ iterations. We have done some other configurations with a finer time step, but the quality of the results are comparable to the ones shown in the letter.

In this supplement we also consider other values of $N$ (in the letter we had considered the $11\times 11$ case only).

\paragraph{-- Computing autocorrelation functions. }

In the letter we state that we compute the autocorrelation functions
\begin{equation}
f_{\cal O}(a)= \langle {\cal O}(t) {\cal O}^\dagger(t+a)\rangle\,,\nonumber
\end{equation}
 where ${\cal O}(t)$ is some classical gauge invariant observable. These are averages over time, so the variable $t$ is averaged over. These are good measurements of the dynamics at equilibrium. The initial time cutoff is controlled by asking when has the system thermalized. In the letter in Fig.~$2$ we have shown that the size of the system reaches equilibrium values at some 
$t$ close to $500$ (in iterations that are saved). We wait until machine time is 5000 before we start averaging. Our simulation runs for  20000 stored iterations.  We get the autocorrelation functions from computing the Fourier transform of the power spectrum of the last 15000 saved iterations. These depend on the time displacement $a$ and on the observable.

Due to rotational symmetry of the configurations there are various autocorrelation functions that are the same. We can use this to get a measurement of statistical error bars with a sample of 3 from a  single run. We normalize these to the standard deviation of fluctuations, so the plot depicts the quantities
\begin{equation}
f_{\cal O}(a)/f_{\cal O}(0)\nonumber
\end{equation}
and we use the average value $f_{\cal O}(0)$ to do this. Considering we have various data sets, they display some small variation in the size of fluctuations. This is expected.


\subsection{Additional plots}

\paragraph{ -- Varying the initial velocity of the single eigenvalue.}

\begin{figure*}[ht]
\begin{center}
\begin{tabular}{cc}
\includegraphics[scale=.45]{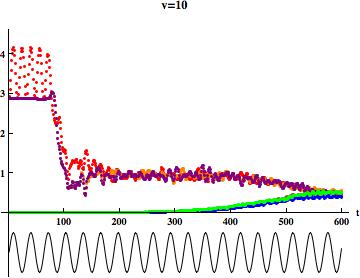}\qquad\qquad
& \qquad\qquad
\includegraphics[scale=.45]{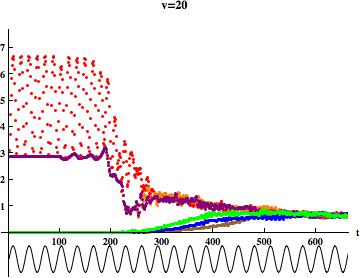}
\\
\includegraphics[scale=.45]{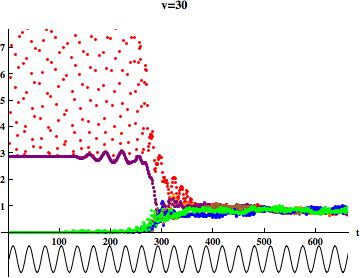}\qquad\qquad
& \qquad\qquad
\includegraphics[scale=.45]{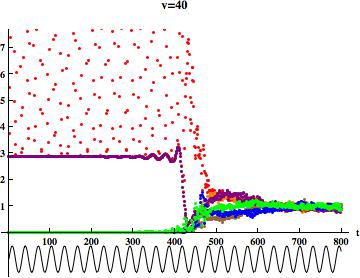}
\\
\includegraphics[scale=.45]{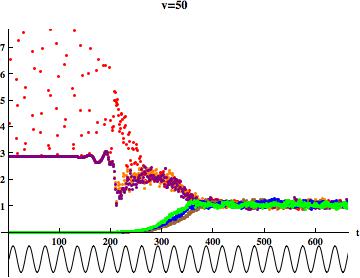}\qquad\qquad
& \qquad\qquad
\includegraphics[scale=.45]{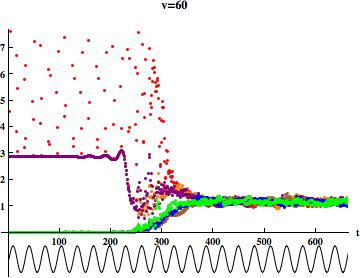}
\\
\includegraphics[scale=.45]{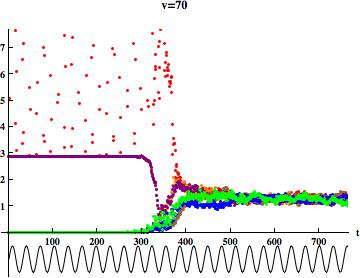}\qquad\qquad
&\qquad\qquad
\includegraphics[scale=.45]{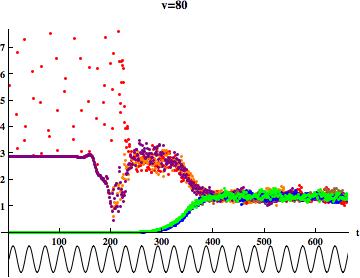}
\\
\includegraphics[scale=.45]{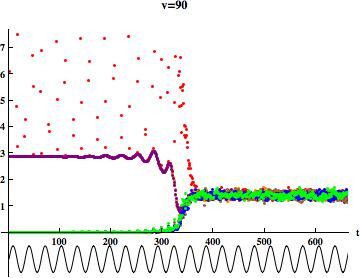}\qquad\qquad
&\qquad\qquad
\includegraphics[scale=.45]{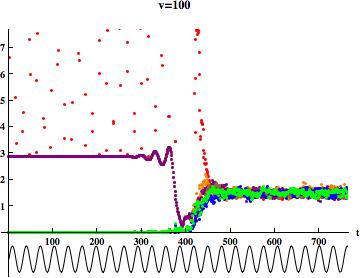}
\end{tabular}
\caption{\small Simulations with different values of the initial velocity $v$. The time to thermalization is found to be roughly independent of $v$.
\label{v}}
 \end{center}
 \end{figure*}

In Fig.~\ref{v}  of this supplement we plot the standard deviations of the eigenvalues of the matrices $X^i$ and $Y^a$ as in Fig.~2 of the letter, but for different initial values of the velocity $v$. We go from $v=10$ to $v=100$ in steps of 10. From these plots we see that the confluence time at which all the standard deviations of the various variables converge to the same value is roughly independent of $v$. This confluence time is approximately given by 15 - 18 periods of the internal clock (the black curve at the bottom of the plots).

The rest of the initial conditions are as follows:
\begin{equation}
N=10+1\,, ~~ \tilde\hbar = 10^{-6}\,,~~ \Delta t=0.0008\,,~~  N_{cycles}=250\,.\nonumber
\label{parameters}
\end{equation}
The variable $N_{cycles}$ dictates how long we wait before writing new configurations to a file. 
Notice the relatively high value for $N_{cycles}$. We had selected this value because we wanted to look at long simulations without generating huge dump files, but of course this makes the curves look more discontinuous than what we had in the letter (especially for $X^0$, the red curve in the plots). 

These also were obtained at a smaller value of $\hbar$ than what is reported on the letter. We will have further comments on this later on in this supplement. These were generated 
in a previous iteration of the code where the definition of $\hbar$ is slightly different, thus the tilde on top of it. The main reason for the new definition was to make it possible to fix the quantity $\hbar$ to be able to compare different values of $N$ homogeneously.

\paragraph{-- Varying the size of the matrices.}

To show that our results are fairly independent of $N$, we now include plots for $N=20+1$ similar to the ones contained in the letter for the $N=10+1$ case. 
\begin{figure*}[ht]
\includegraphics[width=3.3 in]{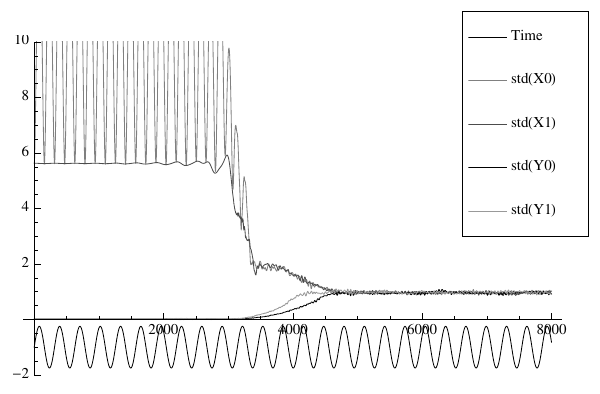} \qquad
\includegraphics[width=3.3 in]{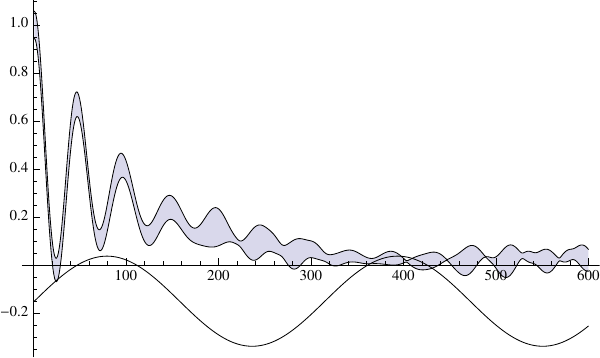} 	
\caption{Left: simulation of size of configuration in various directions as measured with standard deviation of eigenvalues at time $t$. We include our clock measure (black curve at the bottom). Right: normalized autocorrelation functions of $\tr[(X_1+iX_2)^2]$ and comparison to the clock. The width of the band indicates our statistical uncertainty from few similar sequences related to each other by rotations. } 
\label{fig:autocorr1suppl}
\end{figure*}

The figures shown in Fig.~\ref{fig:autocorr1suppl} are calculated with $N=20+1$, $v=60$, $\hbar= 10^{-4}$, $\delta t= 10^{-4}$. We record the configuration every $20$ iterations. The cutoff for thermalization is taken at saved iteration $5000$, immediately when the size lines coincide. This simulation collected 25000 iterations from the beginning.  The plots here are generated in the same way that the same plots used in the letter in Figs. 2 and 4 (the top one) we generated. They are the plots for the new configuration. The plots are qualitatively similar to the ones included in the letter. A detailed comparison between different $N$ is something we are looking into in order to characterize the system better. Notice that the autocorrelation functions have a different natural oscillation time relative to the clock.

\begin{figure*}[ht]
\begin{center}
\begin{tabular}{cc}
\includegraphics[scale=.6]{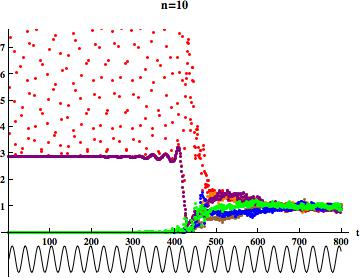}\qquad\qquad
& \qquad\qquad
\includegraphics[scale=.6]{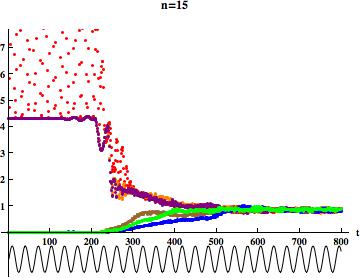}
\\
\\
\includegraphics[scale=.6]{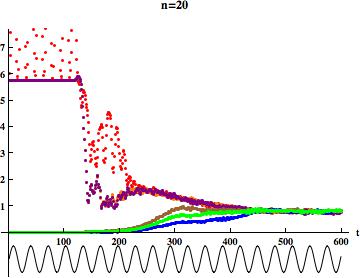}\qquad\qquad
&\qquad\qquad
\includegraphics[scale=.6]{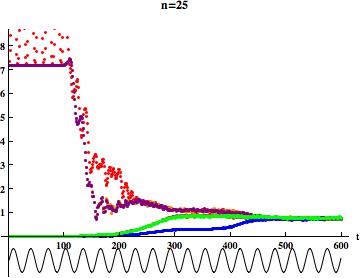}
\\
\end{tabular}
\caption{\small Simulations with different values of the matrix size $N=n+1$. The initial velocity is fixed at $v=40$ and the rest of the parameters are as in (\ref{parameters}). The time to thermalization is found to be roughly independent of $n$ (at $t\sim 500$).
\label{n}}
 \end{center}
 \end{figure*}

In Fig.~\ref{n} we also include plots equivalent to the ones in Fig.~\ref{v}, this time keeping $v=40$ fixed but changing $N$. Again, we see that the confluence times are quite independent of $N$.


\subsection{Thermalization times}

A final technical note that we include here is regarding the notion of thermalization times. There are three notions of thermalization times that we can consider. 

First, we can ask what is the time it takes for a small quantum fluctuation (adding a particular quantum bit) to scramble so that we can not recover that bit any longer as a function of the size of the system. This is the question that is asked in the paper by Sekino and Susskind \cite{SS}, where it is argued that black holes are fast scramblers. This question is for a near equilibrium situation.
 Our code does not provide yet for a way to add additional fluctuations after a black hole is formed, so we can not address this question directly. Also, we have not explored how to change the size of the dual black holes systematically: we need to change $N$ keeping something fixed as we vary $N$. It is not clear from the paper \cite{SS} what is the correct way to implement this idea in our simulations. 

Secondly, there is an intrinsic notion of scrambling near equilibrium as provided by the fluctuation dissipation theorem. Understanding how the autocorrelations die gives us a classical notion of how long it takes for a fluctuation in a particular channel to die. As described in the letter in Fig. 4, and in the supplement in Fig. \ref{fig:autocorr1suppl}, the autocorrelation functions both decay and oscillate. The decay of the signal as measured from $a=0$ to the next peak in the oscillation is about $0.5-0.6$ and within about 3 oscillations it has decayed to about less than a third of the size. This means that the autocorrelation function dies off as quickly as it oscillates and can therefore be considered a fast scrambler from this point of view. This ratio of times does not seem to depend too strongly on the size of the system.

Third, there is the time to thermalization from a particular initial condition. Since the only place where $\hbar$ is used in the letter is in the details of the initial conditions, we can ask what is the $\hbar$ dependence of the time to thermalization fixing everything else. When $\hbar$ is zero, the configuration is in periodic motion forever and it does not thermalize. When $\hbar$ is small, the off-diagonal fluctuations are tiny and we can solve the equations for these modes in the linearized regime. This was done in detail in the previous work by some of the authors~\cite{Berenstein:2010bi}. The fluctuations are either stable or unstable. The unstable modes grow exponentially between oscillations until back reaction of the nonlinearities kicks in. 
The nonlinearities can be understood as fluctuations reaching a particular size. We can call this size of fluctuation the saturation size and the time at which the fluctuations get to this size the saturation time. 
Before this, the growth is exponential, so the time for this growth depends on the size of the initial fluctuation logarithmically. This is $t\propto \log( A_{sat}/A_{in})$. Since 
$A_{in} \propto \sqrt{\hbar}$, the time to saturation grows logarithmically in $\hbar$. Evidence for this can be seen by comparing the second figure in this supplement with respect to the corresponding Fig. 2 in the letter. Notice also that after the $X$ modes shrink, it takes longer for the $Y$ modes to saturate in the plot in this supplement relative to the one appearing in the letter. This is because the $Y$ modes grow slower in the initial stage, so their amplitude is small after the first collapse. 

Notice that if we keep $X$ random, but $Y$ infinitesimal in the equations of motion, the $Y$ motion is linear in $Y$. So again, if the initial size after the first collapse of $X$ is controlled by $\hbar$, the time to saturation of the Y modes is logarithmic in $\hbar$.

The plots give evidence for this behavior.  Obviously, if we increase $\hbar$ we shorten the time to thermalization from the initial condition. Care has to be taken on what exactly is an equilibrium configuration, so this time depends on the systematics of how we declare that a particular configuration has reached equilibrium.

\end{document}